\documentclass[a4paper,11pt]{article}
\usepackage{pos}
\usepackage{graphicx}  
\usepackage{caption}
\usepackage{subcaption}
\usepackage{moresize}

\title{The isospin-violating part of the hadronic vacuum polarisation}

\author*[a]{Dominik~Erb}
\author[b]{Antoine~Gerardin}
\author[a,c]{Harvey~B.~Meyer}
\author[a]{Julian~Parrino}
\author[e]{Vladimir~Pascalutsa }
\author[e]{Volodymyr~Biloshytskyi}

\affiliation[a]{PRISMA$^+$ Cluster of Excellence \& Institut f\"ur Kernphysik,
Johannes Gutenberg-Universit\"at Mainz,
D-55099 Mainz, Germany}

\affiliation[b]{Aix-Marseille Universit\'e  , Universit\'e de Toulon, CNRS, CPT, Marseille, France}

\affiliation[c]{Helmholtz~Institut~Mainz,
Staudingerweg 18, D-55128 Mainz, Germany}

\affiliation[e]{Institut f\"ur Kernphysik,
Johannes Gutenberg-Universit\"at Mainz,
D-55099 Mainz, Germany}

\emailAdd{domerb@uni-mainz.de}

\abstract{We present our calculation of the isospin-violating part of the hadronic vacuum polarisation (HVP) contribution to muon $(g-2)$ in lattice QCD at the $SU(3)_{\mathrm{f}}$ symmetric point. The computation of the contributing fully connected diagrams with one internal photon as well as the computation of the only (mass) counterterm are shown. The latter is determined from the charged-neutral kaon mass splitting. We employ coordinate-space methods and a photon propagator which is regulated \`a la Pauli-Villars with a cutoff scale $\Lambda$ well below the lattice cutoff. This regularization makes it possible for us to do crosschecks of individual contributions with calculations in the continuum. Our continuum extrapolated results show little to no dependence on $\Lambda$. This makes our final limit $\Lambda \rightarrow \infty$ straightforward. }

\FullConference{The 41st International Symposium on Lattice Field Theory (LATTICE2024)\\
 28 July - 3 August 2024\\
Liverpool, UK\\}

\begin{document}
\maketitle

\section{Introduction}
With the recent muon $(g-2)$ measurement at Fermilab \cite{Muong-2:2021ojo, Muong-2:2023cdq} the global experimental average reached a 5.2 $\sigma$ discrepancy with the 2020 estimate of the Theory Initiative \cite{Aoyama:2020ynm}. Additionally, the uncertainty of the Standard-Model value now exceeds the one put forth by the experimental effort. While the main contribution to the value of the former is due to QED effects, the main contribution to the uncertainty is due to hadronic effects. The value and uncertainty of these hadronic effects themselves are dominated by the hadronic vacuum polarisation (HVP), which enters at $O(\alpha^2)$.\\
One way of calculating the HVP from first principles is using Lattice QCD (LQCD). 
While LQCD simulations usually do not include different masses and QED effects for the light quarks, one has to account for these differences to reach sub-percent precision.
We will focus here on the QED effects which enter at $O(\alpha^3)$ and can be accounted for by considering the next-to-leading order expansion of the QCD vector-vector correlator. These calculations bear resemblance to the hadronic light-by-light contribution (HLbL), which is the secound largest contribution to the hadronic effects of the muon $(g-2)$. However, while the HLbL contribution is UV-finite, the additional photon leads to divergences in the HVP calculation. To address this issue, we will use the (double) Pauli-Villars (PV) regularisation for the photon propagator, which was proposed in Ref. \cite{Biloshytskyi:2022ets}. For the calculation of the HVP on the lattice, we will use the covariant coordinate-space method \cite{Meyer:2017hjv}.\\
In this work, we will present our calculation of the isospin-violating part of the HVP at the $SU(3)$ flavour symmetric point. For that, we will first give an overview over our formalism in section \ref{sec::CCS}. While there are many diagrams which contribute at $O(\alpha^3)$ to the HVP, we will mostly focus on the fully connected diagrams. Their calculation and results are discussed in section \ref{sec::Connected}. With the additional scale introduced by the inclusion of QED effects, the addition of a (mass) counterterm becomes necessary. Section \ref{sec::Counterterm} gives a brief overview over the contributions to this counterterm, focusing on the charged-neutral kaon mass splitting, shortened to kaon mass splitting, which also includes the PV-regulated photon propagator. Lastly, the HVP calculation and the counterterm are combined and extrapolated to the continuum in section \ref{sec::Extra}. The section concludes with a discussion on the dependence of the continuum extrapolated values on the PV mass. 

\section{Covariant coordinate-space method \label{sec::CCS}}
In order to calculate the isospin-violating part of the HVP, we employ the covariant coordinate-space (CCS) method \cite{Meyer:2017hjv}. This method is a proven alternative \cite{Chao:2022ycy} to the time-momentum representation (TMR), which is usually used to calculate this quantity on the lattice. In the CCS method the HVP contribution to the muon $(g-2)$ is obtained from a space-time integral weighted by a kernel. There is some freedom in choosing the kernel, which is one advantage of the CCS method. In this work, we use the traceless (TL) version of the kernel from \cite{Chao:2022ycy}, which takes the form 
\begin{align}
    H_{\lambda \sigma}^{TL}(z)= \left(-\delta_{\lambda \sigma} + 4\frac{z_\lambda z_\sigma }{z^2}\right)\mathcal{H}_2(|z|). \label{equ::Kernel}
\end{align}
The exact form of the scalar weight function can be found in \cite{Meyer:2017hjv}. 
The integral we use to calculate the NLO isospin-violating part is given by equation \eqref{equ::a_mu}. To obtain this representation, the QCD path integral was expanded in the electromagnetic coupling. From this we also get a photon propagator. It is important to note that the bare electromagnetic correction to HVP is not UV finite, in contrast to the HLbL contribution. We therefore use a (doubly) Pauli-Villars regulated photon propagator \cite{Biloshytskyi:2022ets}. Its position space representation is given by equation \eqref{equ::Photon} were $K_1(x)$ is the modified Bessel function of the second kind. The additional scale introduced by QED necessitates the introduction of a counterterm which is also included in equation \eqref{equ::a_mu}. The counterterm we employ here can be seen in equation \eqref{equ::Counterterm}. The first factor comes form the determination of the light-quark mass splitting. It consists of the physical charged-neutral kaon mass splitting  $\Delta m_K^{phys} = -3.934$ MeV (from the PDG \cite{PDG}), the kaon mass splitting $\Delta m_K^{em}$ calculated on the lattice using the PV-regulated photon propagator and a mass insertion into the kaon propagator. This combination then gets multiplied by the derivative of the leading order HVP contribution with respect to the light-quark mass difference:
\begin{align}
    a_\mu^{\text{HVP}, \text{NLO}, 38} &=-\frac{e^2}{2}\int_{z,x,y} \ H^{TL}_{\lambda \sigma}(z) \delta_{\mu \nu} [\mathcal{G}(x-y)]_\Lambda \langle j^3_\lambda(z) j^{em}_\mu(x) j^{em}_\nu(y) j^8_\sigma (0) \rangle +CT \label{equ::a_mu}\\
j^3_\mu=\frac{1}{2}&\left( \Bar{u}\gamma_\mu u - \Bar{d}\gamma_\mu d\right), \quad j^8_\mu=\frac{1}{6}\left( \Bar{u}\gamma_\mu u + \Bar{d}\gamma_\mu d - 2\Bar{s}\gamma_\mu s\right), \quad j^{em}_\mu= j^3_\mu + j^8_\mu\\
    CT&=- \frac{\Delta m_K^{em}-\Delta m_K^{phys}}{ \langle K^+| \Bar{u}u-\Bar{d}d | K^+ \rangle } \  \frac{\partial a_\mu ^{\text{HVP}, 38}}{\partial(m_u-m_d)} \biggr\rvert_{m_u+m_d, \ m_s,\  g_0}  \label{equ::Counterterm}\\
[\mathcal{G}(x)]_\Lambda &= \frac{1}{4\pi^2 |x|^2} - \frac{\Lambda K_1\left(\Lambda \frac{|x|}{\sqrt{2}}\right)}{2\sqrt{2}\pi^2 |x|} + \frac{\Lambda K_1\left(\Lambda |x|\right)}{4\pi^2 |x|}.   \label{equ::Photon}
\end{align}

\section{Calculating the connected contribution \label{sec::Connected}}
\begin{figure}[t]
    \centering
    \begin{subfigure}{0.43\textwidth}
        \includegraphics[width=\textwidth]{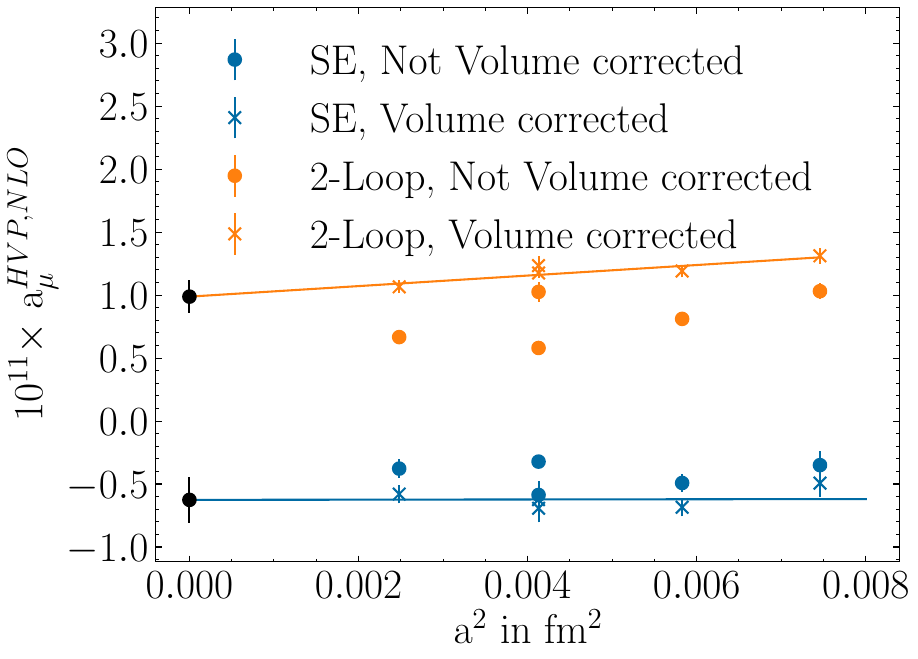}
        \caption{Self-energy and 2-Loop}
    \end{subfigure}
    \begin{subfigure}{0.43\textwidth}
        \includegraphics[width=\textwidth]{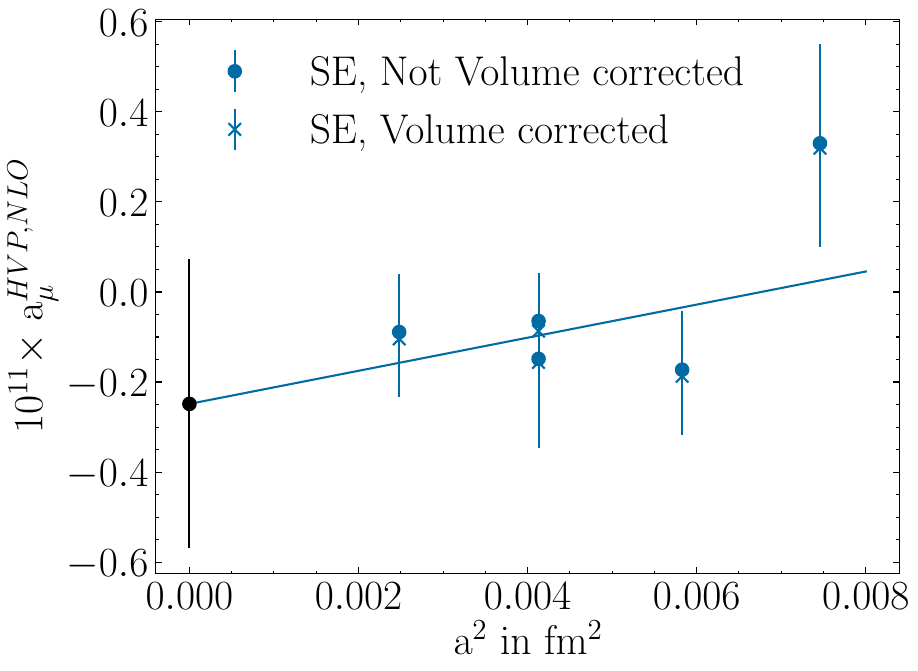}
        \caption{Total=2$\times$SE+2Loop}
    \end{subfigure}
\caption{Continuum extrapolation of the self-energy and 2-loop parts on the left side as well as the total value on the right side for the QCD ensembles. The Pauli-Villars mass is set to $\Lambda = 16\, m_\mu$. 
Equation \eqref{equ::Fit_Func} is used as an ansatz for the fit.
The dots are the data from the lattices, while the crosses are the same data, but with the volume term of the fit function subtracted without adjusting the error bars. The straight lines are the fits to these volume-corrected points. The black dots are the results of the continuum extrapolation.}
\label{fig::Conn_QCD_Cont_Extra}
\end{figure}

The ensembles used in these proceedings were generated as part of the CLS (Coordinated Lattice Simulations) initiative. They were obtained using $N_f=2+1$ dynamical flavors of non-perturbatively $O(a)$ improved Wilson-Clover quarks and the tree-level $O(a^2)$ improved Lüscher-Weisz gauge action. The properties of these ensembles are listed in table \ref{tab::LatSet_Properties}. All of them are at the $SU(3)$ flavour symmetric point and have a pion and kaon mass of around 416 MeV. Two of the ensembles have the same lattice spacing, but different volumes in order to get a better handle on finite-size effects. Lastly, one of the ensembles, B450, has periodic boundary conditions in time while the other ones have open boundary conditions in time.
\begin{table}[b]
    \centering

    \begin{tabular}{ c c c c c c c c}
    \hline
        Id & $\beta$ & $(\frac{L}{a})^3 \times \frac{T}{a}$ & a [fm] & $m_\pi$ [MeV] &  $m_\pi L$ & L[fm] & $\hat{Z}_V$\\ \hline
        H101 & 3.4 & $32^3 \times 96$ & 0.08636 & 416(4) & 5.8 & 2.8 & 0.71562 \\ \hline 
        B450 & 3.46 & $32^3 \times 64$ & 0.07634 & 415(4) & 5.1 & 2.4 & 0.72647 \\ \hline
        H200 & 3.55 & $32^3 \times 96$ & 0.06426 & 416(5) & 4.3 & 2.1 & 0.74028 \\
        N202 &  & $48^3 \times 128$ &  & 412(5) & 6.4 & 3.1 & \\ \hline
        N300 & 3.7 & $48^3 \times 128$ & 0.04981 & 419(4) & 5.1 & 2.4 & 0.75909 \\ \hline
    \end{tabular}
    \caption{Parameters of the used CLS ensembles. The lattice spacing in physical units was extracted from \cite{Bruno:2016plf}. The pion mass values were taken from \cite{Ce:2022kxy}. We used \cite{Gerardin_2019} to extract $\hat{Z}_V$ for each value of $\beta$. The B450 ensemble has periodic boundary conditions in time, while all other ensembles have open boundary conditions in time.}
    \label{tab::LatSet_Properties}
\end{table}
\begin{figure}[ht]
	\centering
	\includegraphics[width=0.5\textwidth]{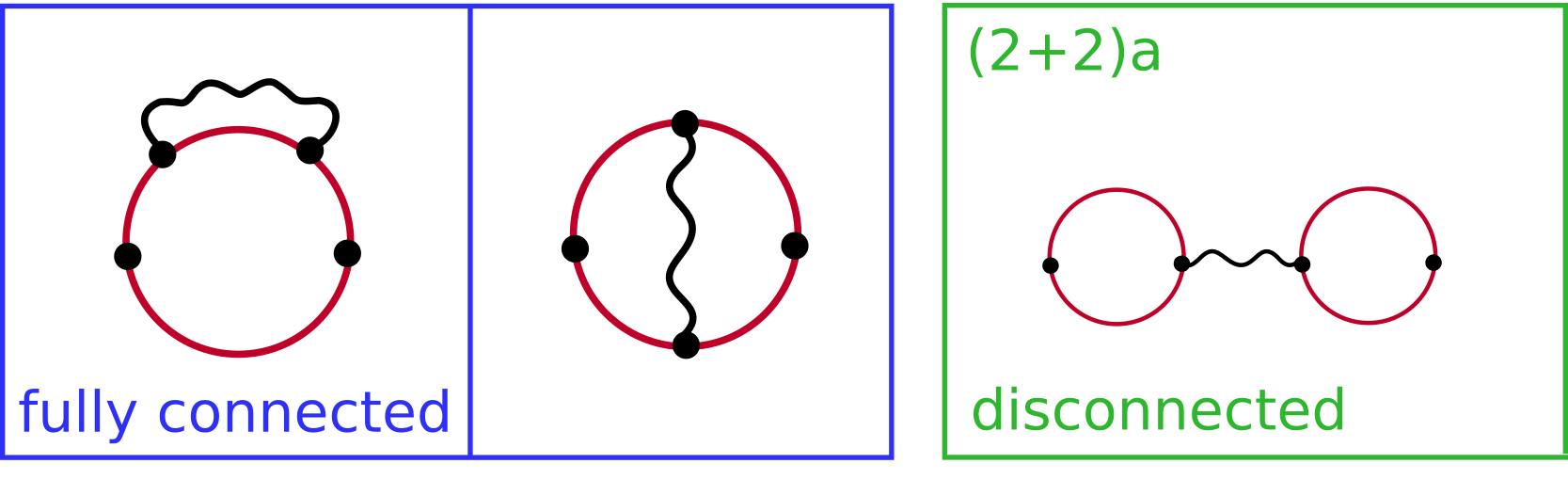}
	\caption{Diagrams that contribute to $\langle j^3_\lambda(z) j^{em}_\mu(x) j^{em}_\nu(y) j^8_\sigma (0) \rangle$ at the $SU(3)_{\mathrm{f}}$ symmetric point.}
    \label{fig::Master_feynman}
\end{figure}\\
At the $SU(3)_{\mathrm{f}}$ symmetric point, there are three Feynman diagrams which contribute to the QCD four-point function $\langle j^3_\lambda(z) j^{em}_\mu(x) j^{em}_\nu(y) j^8_\sigma (0) \rangle$. These diagrams can be seen in figure \ref{fig::Master_feynman}. In the following, we will discuss the calculation of the fully connected diagrams. The result of the (2+2)a disconnected diagram will be added later on in section \ref{sec::Extra}, already in the continuum. \\
The fully connected diagram on the left will be referred to as the self-energy diagram, while the one on the right will be called the 2-loop diagram. The corresponding expectation values over gauge configurations, $\langle ... \rangle_U$, with only local vector currents written in terms of propagators for each of these diagrams are given in equations \eqref{equ::Exp_SE} and \eqref{equ::Exp_2L}, respectively. 
\begin{align}
    C_{SE}(x, y, z)&=  -2 \, \text{Re} \langle \text{Tr} [ S(0, x) \gamma_\nu S(x, y) \gamma_\mu S(y, z) \gamma_\lambda S(z, 0) \gamma_\sigma] \rangle_U \label{equ::Exp_SE} \\
    C_{2L}(x, y, z)&=  -2 \, \text{Re} \langle \text{Tr} [ S(0, y) \gamma_\mu S(y, z) \gamma_\lambda S(z, x) \gamma_\nu S(x, 0) \gamma_\sigma] \rangle_U \label{equ::Exp_2L} \\
    a_{\mu, conn.}^{\text{HVP}, \text{NLO}, 38}&= -\mathcal{Q} \mathcal{Z} \frac{e^2} {2}\int_{z,x,y} \ H^{TL}_{\lambda \sigma}(z) \delta_{\mu \nu} [\mathcal{G}(x-y)]_\Lambda (2\, C_{SE}(x, y, z) + C_{2L}(x, y, z)) \label{equ::Conn_a_mu_1}
\end{align}\\
The contribution from just these two connected diagrams is written in equation \eqref{equ::Conn_a_mu_1}. The charge factor $\mathcal{Q}$ is here equal to 1/36. The local vector currents need to be renormalized. For that we use \cite{Gerardin_2019} to calculate the renomalization factor $\hat{Z}_V$ for each value of the inverse coupling. $\mathcal{Z}$ summarises the needed renormalization factors. Going forward we will use local ($l$) as well as conserved ($c$) vector currents for one of the internal ($X$) and one of the external ($Z$) vertices. One last important thing to note about equation \eqref{equ::Conn_a_mu_1} is the factor of two in front of the self-energy part. This comes from the fact that the photon line in figure \ref{fig::Master_feynman} can be on the upper as well as the lower propagator, which means there are in total two diagrams contributing to this part.
In order to confirm the viability of our methodology, we first present results from ensembles without the strong interaction, i.e. where the link variables $U$ are set to unity. The continuum extrapolated result from such ensembles can be crosschecked with continuum calculations because of our use of the PV-regularization. Our extrapolated results for the four different discretisations can be seen in table \ref{tab::Cont_Gluonless}. A detailed description of the continuum calculation can be found in \cite{Biloshytskyi:2022ets}. The continuum result is $-7.50 \times 10^{-11}$.
The values shown in the table are very close to the continuum results for all discretisations, except for the one with no conserved vector currents, due to its difficult extrapolation.
Since the continuum extrapolation of the XcZl and XcZc discretisations were especially flat with regards to the lattice spacing, we choose these two for our calculations on the CLS ensembles. 
\begin{table}[h]
    \begin{subtable}[h]{0.45\textwidth}
        \centering
	\begin{tabular}{c c c c c}
		\hline
		 & XlZl & XcZl & XlZc & XcZc \\ \hline
		Total & $-6.90$ & $-7.36$ & $-7.44$ & $-7.56$ \\ \hline
		2-Loop & $28.62$ & $28.32$ & $28.42$ & $28.18$ \\ \hline
		Self-Energy & $-17.76$ & $-17.84$ & $-17.93$ & $-17.87$ \\ \hline
			
	\end{tabular}
       \caption{Gluonless at $\Lambda= 3\, m_\mu$.}
       \label{tab::Cont_Gluonless}
    \end{subtable}
    \hfill
    \begin{subtable}[h]{0.45\textwidth}
        \centering
			
	\begin{tabular}{c c c}
		\hline
		 & XcZl & XcZc \\ \hline
		Total & $-0.25(33)$ & $-0.24(31)$ \\ \hline
		2-Loop & $0.99(13)$ & $0.86(13)$ \\ \hline
		Self-Energy & $-0.63(19)$ & $-0.55(18)$ \\ \hline
			
	\end{tabular}
        \caption{With gluons at $\Lambda= 16\, m_\mu$.}
        \label{tab::Cont_With_Gluons}
     \end{subtable}
     \caption{Results of the continuum extrapolation for (a) the gluonless ensembles and (b) the CLS ensembles for different discretisations. }
     \label{tab::Cont}
\end{table}\\
The continuum extrapolation for the CLS ensembles can be seen in figure \ref{fig::Conn_QCD_Cont_Extra}. The used fit function includes a volume term to correct for finite-size effects,
\begin{align}
    f_{fit}(\textbf{a}, m_\pi L)=b + c \, \textbf{a}^2 + d \, e^{-\frac{m_\pi L}{2}}. \label{equ::Fit_Func}
\end{align}
The plots show the direct lattice data as round points as well as the same data but with the volume term corrected as crosses. It can be seen that for the two parts separately the volume correction is important, but when added together it is not statistically significant. \\
The black dots show the results of the continuum extrapolation, with the numerical value written in table \ref{tab::Cont_With_Gluons}. It is clearly visible that the self-energy part and the 2-loop part are non-zero, even within error, but in the sum there is a huge cancellation, which results in the total contribution being equal to zero within error. 

\section{Calculating the counterterm \label{sec::Counterterm}}
\begin{figure}[ht]
		\centering
		\begin{subfigure}{0.4\textwidth}
			\includegraphics[width=\textwidth]{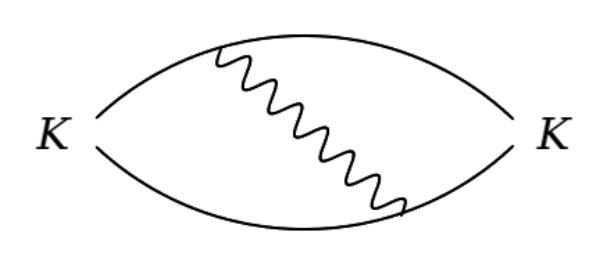}
			\caption{K1 diagram}
		\end{subfigure}
		\begin{subfigure}{0.4\textwidth}
			\includegraphics[width=\textwidth]{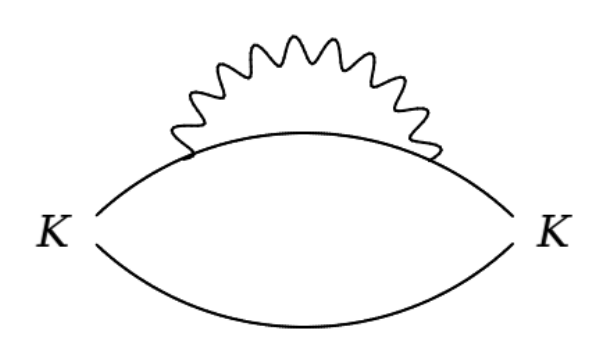}
			\caption{K2 diagram}
            \label{fig::Kaon_MS_Feynman_asymm}
   
		\end{subfigure}
  \caption{The Feynman diagrams of the leading order contributions to the electromagnetic mass splitting of the kaon. At the $SU(3)_{\mathrm{f}}$ symmetric point only these two diagrams contribute to the mass splitting \cite{de_Divitiis_2013}.}
  \label{fig::Kaon_MS_Feynman}

\end{figure}
\begin{figure}[ht]
		\centering
		\begin{subfigure}[t]{0.43\textwidth}
			\includegraphics[width=\textwidth]{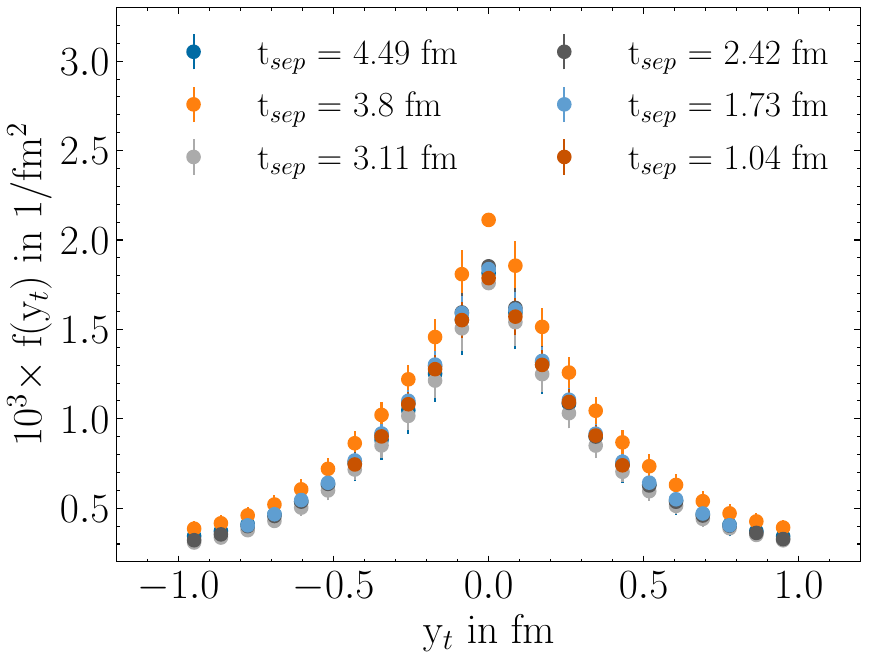}
			\caption{Data restricted to $y_t<1$ fm.}  \label{fig::Kaon_MS_Method_a}
		\end{subfigure}		
        \begin{subfigure}[t]{0.43\textwidth}
			\includegraphics[width=\textwidth]{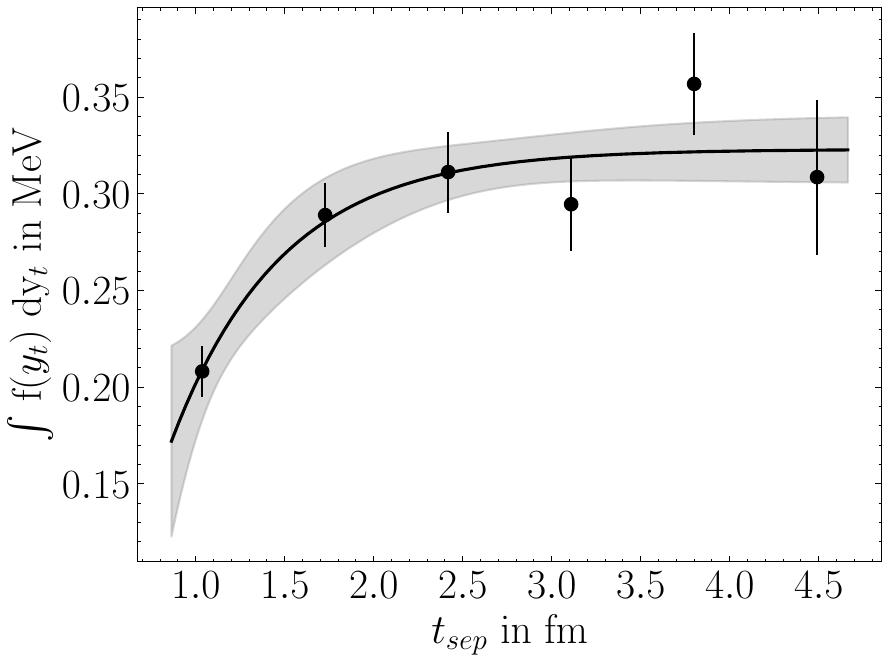}
			\caption{Extrapolation of restricted data.}  \label{fig::Kaon_MS_Method_b}
		\end{subfigure}
  \caption{The methodology discussed in section \ref{sec::Counterterm} for extracting the short distance contribution of the kaon mass splitting. The calculations shown here were done on the H101 ensemble.}
  \label{fig::Kaon_MS_Method}
\end{figure}
\begin{figure}[ht]
		\centering
		\begin{subfigure}[t]{0.45\textwidth}
			\includegraphics[width=\textwidth]{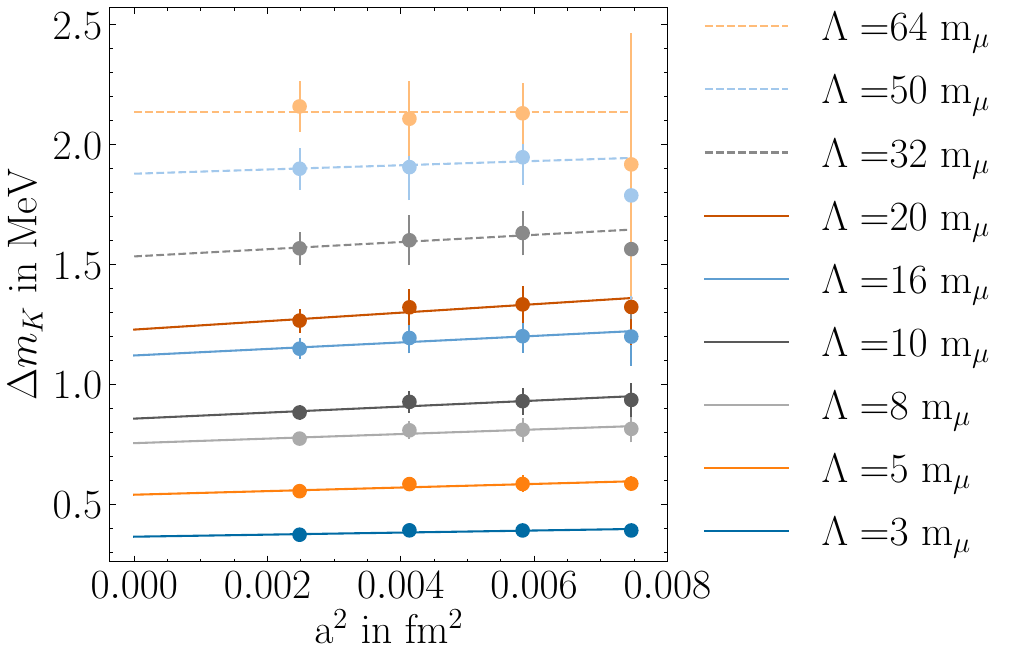}
			\caption{Continuum extrapolation.}  \label{fig::Kaon_MS_Cont_Extra}
		\end{subfigure}		
        \begin{subfigure}[t]{0.40\textwidth}
			\includegraphics[width=\textwidth]{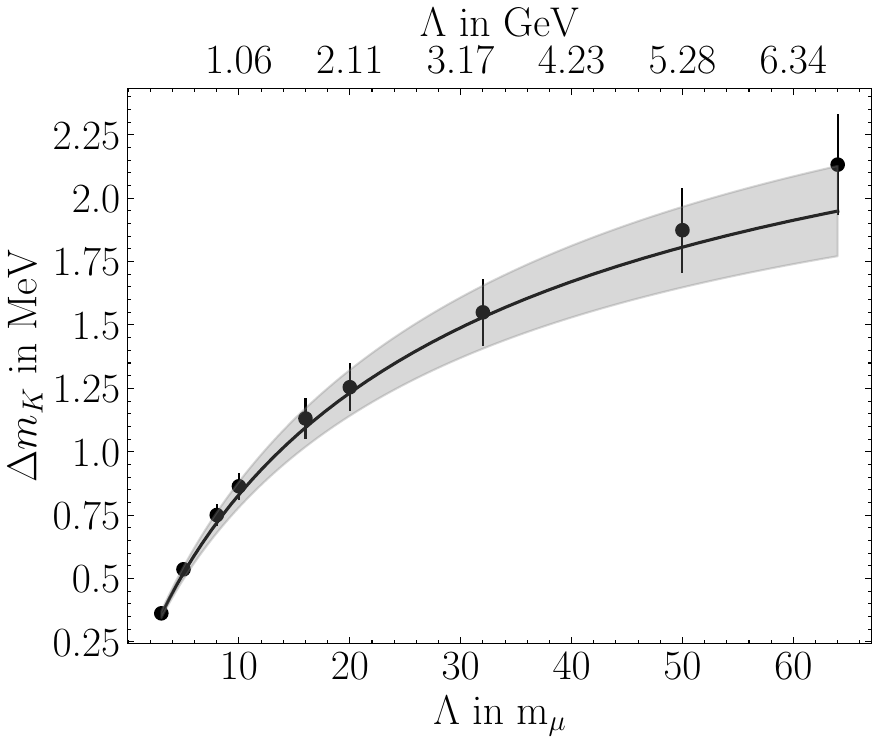}
			\caption{PV-mass behavior.}  \label{fig::Kaon_MS_PV_Extra}
		\end{subfigure}
  \caption{Continuum extrapolation of the kaon mass splitting and PV-mass dependence of the resulting continuum values together with a fit using the fit function in equation \ref{equ::Kaon_MS_fit_func}. The H200 ensemble was excluded from the extrapolation since it showed large volume effects. }
  \label{fig::Kaon_MS_Extra}
\end{figure}

In this section, we discuss the calculation of the counterterm, focusing on the determination of the kaon mass splitting. 
At the $SU(3)_{\mathrm{f}}$ symmetric point there are only two diagrams, which contribute to the bare electromagnetic kaon mass splitting \cite{de_Divitiis_2013}, see figure \ref{fig::Kaon_MS_Feynman}. One interesting observation is that for large distances between the kaon creation and annihilation operators, these diagrams are completely dominated by the elastic contribution, which can be calculated analytically. This leads us naturally to the following methodology. We first calculate the diagrams for different separation times (figure \ref{fig::Kaon_MS_Method_a}). Then we restrict the results to the short distance part. We extrapolate this restricted data to infinite separation times (figure \ref{fig::Kaon_MS_Method_b}) and at the end we add back in the long distance part which we calculated analytically with only the elastic contribution in infinite volume. \\
The continuum extrapolation of nine different PV-mass values is shown in figure \ref{fig::Kaon_MS_Cont_Extra}. In order to check the plausibility of these values, we can examine the large PV-mass behavior. Using an Operator Product Expansion (OPE), similar to the one used in \cite{Biloshytskyi:2022ets} we get the following prediction for $\Lambda \rightarrow \infty$:
\begin{align}
    \Delta m^{em}_K(\Lambda)\approx& \ \frac{3\alpha}{2\pi}\log \left(\frac{\Lambda}{\mu_{IR}}\right) (Q_u^2-Q_d^2)m_l \frac{\partial m_K}{\partial m_l} =\ \mathcal{C} \log \left(\frac{\Lambda}{\mu_{IR}}\right). \label{equ::Kaon_MS_OPE_Prediction}
\end{align}
On the right hand side, all constant factors were summarized in $\mathcal{C}$ which is approximately 0.12 MeV. The extrapolated mass splitting values together with a fit to these values can be seen in figure \ref{fig::Kaon_MS_PV_Extra}. The fit function is based on the OPE prediction and given by 
\begin{align}
     f_{fit}(\Lambda)= a \frac{\Lambda}{\Lambda+d} + \mathcal{C} \log(\frac{\Lambda +b}{b}) \label{equ::Kaon_MS_fit_func}.
\end{align}
It was slightly modified in order to make the function vanish for $\Lambda \rightarrow 0$, which is the behavior we expect, since the photon propagator vanishes in this limit. 
But most importantly the constant factor in front of the logarithm is the same as in equation \eqref{equ::Kaon_MS_OPE_Prediction}. For the fit we took a correlation of 0.9 between the input parameters, based on calculations for each ensemble. The resulting $\chi^2/DOF$ of the fit is 0.77. This indicates the plausibility of the obtained values of the kaon mass splitting. \\
We obtain the second part of the counterterm, the mass insertion into the kaon propagator, using the plateau method. For that we first calculate the matrix element $\langle K^+| \Bar{u}u-\Bar{d}d | K^+ \rangle$ on the CLS ensembles for a separation time between the creation and annihilation of the kaon which is as large as possible. Then we look on the dependence of the matrix element on the insertion time of the mass operator. Lastly, we do a fit to the resulting values with a constant function to get our result. \\
The last part which has to be calculated for the counterterm is the light-quark mass difference derivative of the HVP. 
Here, we use the mass derivatives from the calculation in reference \cite{Ce:2022kxy} obtained in the TMR representation. In order to increase statistics, stochastic wall sources were used for the calculation of this derivative. 

\section{Extrapolation of the total contribution \label{sec::Extra}}
\begin{figure}[ht]
    \centering
    \begin{subfigure}{0.43\textwidth}
        \includegraphics[width=\textwidth]{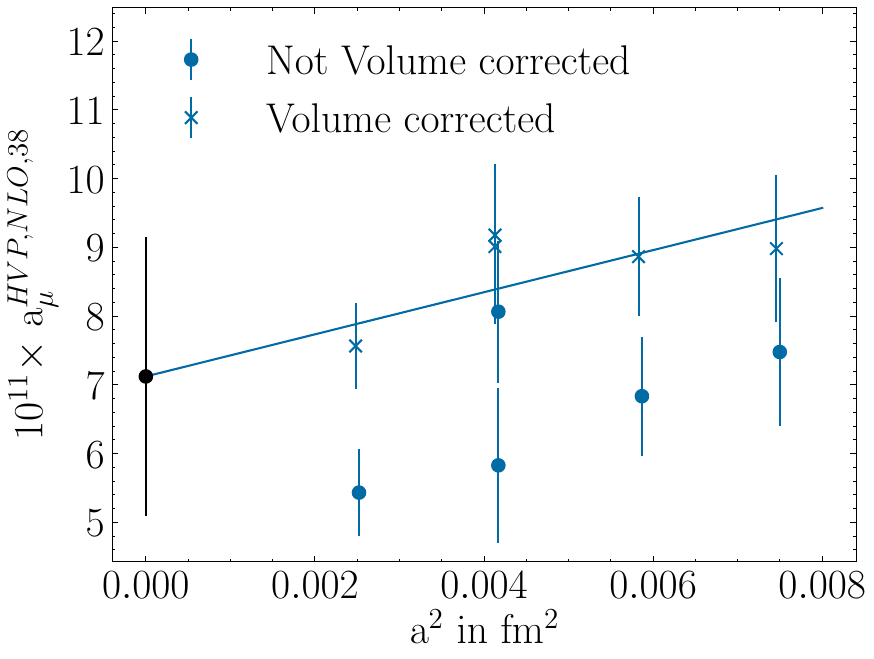}
        \caption{Continuum extrapolation for $\Lambda = 16\, m_\mu$.}
        \label{fig::Total_Extrapolations_Cont}
    \end{subfigure}
    \begin{subfigure}{0.43\textwidth}
        \includegraphics[width=\textwidth]{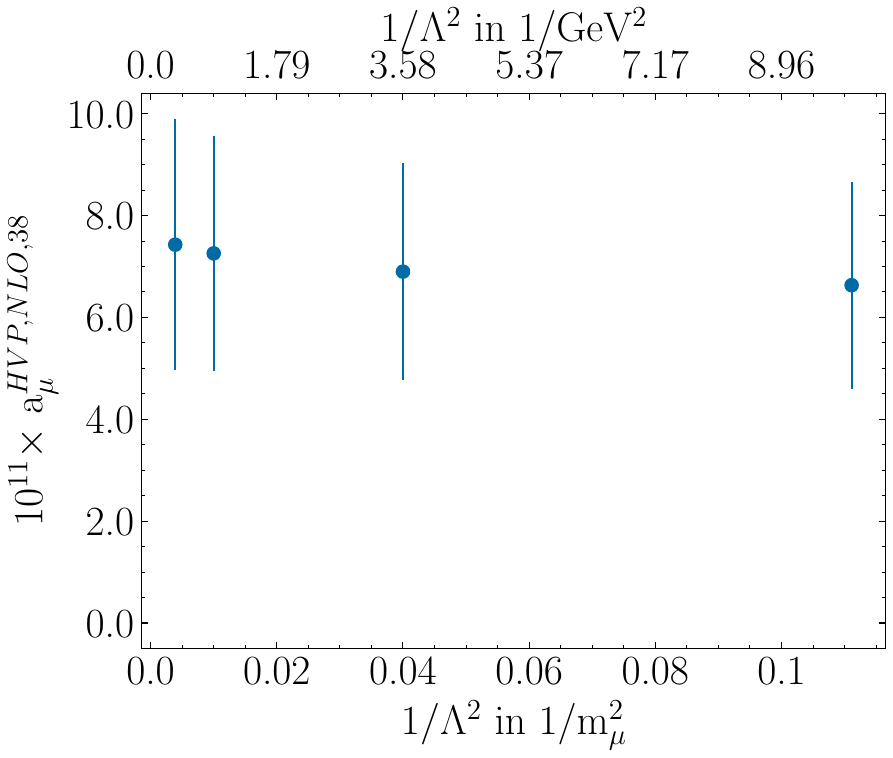}
        \caption{PV-mass extrapolation}
        \label{fig::Total_Extrapolations_PV}
    \end{subfigure}

\caption{Continuum and PV-mass extrapolation of $a_{\mu}^{\text{HVP}, \text{NLO},38}$. Figure (a) is analogous to the plots of figure \ref{fig::Conn_QCD_Cont_Extra} with the same fit function given by equation \ref{equ::Fit_Func} of section \ref{sec::Connected}.  }  
\label{fig::Total_Extrapolations}
\end{figure}
We can now combine the results of the previous sections in order to calculate the total contribution using equation \eqref{equ::a_mu}. First, we perform a continuum extrapolation excluding the (2+2)a disconnected diagram. It was calculated separately without a PV regulator and has a continuum value of $-0.491(82)\times 10^{-11}$. Its charge factor is 1/12. Then we combine the continuum-extrapolated value with the disconnected contribution and look at the dependence on the PV-mass $\Lambda$. \\
The continuum extrapolation for a PV-mass of $16\, m_\mu$ can be seen in figure \ref{fig::Total_Extrapolations_Cont}. The fit function from equation \eqref{equ::Fit_Func} is used, which is the same as the one used in section \ref{sec::Connected}.
The results of the continuum extrapolations combined with the disconnected contribution for the different PV-masses are noted in table \ref{tab::Total_Extrapolated_Vals}. These values are plotted in dependence of the inverse square of $\Lambda$ in figure \ref{fig::Total_Extrapolations_PV}. It can be seen that all the points are equal to one another within error. This means our results have basically no dependence on the PV-mass.\\
One interesting observation is that the values as well as the errors are completely dominated by the counterterm. Taking the N202 ensemble at a PV-mass of $16\, m_\mu$ as an example, the connected contribution has a value of $-0.148 (191)\times 10^{-11}$ while the value of the counterterm is $9.54 (1.24) \times 10^{-11}$. It is expected that this will change when going away from the $SU(3)_{\mathrm{f}}$ symmetric point towards the physical one.
\begin{table}[ht]
	\centering
	\begin{tabular}{c c c c c}
		\hline
		$\Lambda$ in $m_\mu$ & 3 & 5 & 10 & 16 \\ \hline
		$1/\Lambda^2$ in $1/m_\mu^2$ & 0.11 & 0.04 & 0.01 & 0.004 \\ \hline
		  $10^{11}\times a_\mu ^{\text{HVP}, \text{NLO}, 38}(\Lambda)$ & $6.63 \pm 2.04$ & $6.90 \pm 2.13$ & $7.26 \pm 2.32$ & $7.43 \pm 2.47$\\ \hline
	\end{tabular}
 \caption{Continuum extrapolated values of the total contribution for the different PV-mass values.}
 \label{tab::Total_Extrapolated_Vals}
\end{table}

\section{Conclusion \label{sec::Conclusion}}
We have successfully calculated the leading order isospin-violating part of the HVP contribution to muon $(g-2)$ at the $SU(3)_{\mathrm{f}}$ symmetric point using a PV regulated photon propagator. Because of this regularisation we were able to crosscheck our methodology for calculating the connected contribution with continuum calculations by using gluonless ensembles. 
On the CLS ensembles we got comparatively small results for the connected contribution e.g. $-0.25(33)\times 10^{-11}$ for a PV-mass of $16\, m_\mu$. During the calculation of the counterterm we saw the expected logarithmic behavior of the electromagnetic kaon mass splitting for large PV-masses. After doing the combined continuum extrapolation our results show little to no dependence on the PV-mass. By averaging the results with the largest two PV-masses and taking a conservative correlation coefficient estimate of unity we get a final result of $7.3(2.1) \times 10^{-11}$. \\

\fontsize{10}{10} \selectfont
\acknowledgments We thank Simon Kuberski for providing the data for the calculation of the HVP light-quark mass derivative. We also thank Franziska Hagelstein for an ongoing collaboration on computing
electromagnetic corrections to hadronic vacuum polarization. We acknowledge the support of Deutsche Forschungsgemeinschaft (DFG) through the research unit FOR 5327 “Photon-photon interactions in the Standard Model and beyond exploiting the discovery potential from MESA to the LHC” (grant 458854507), and through the Cluster of Excellence “Precision Physics, Fundamental Interactions and Structure of Matter” (PRISMA+ EXC 2118/1) funded within the German Excellence Strategy (project ID 39083149). Calculations for this project were partly performed on the HPC clusters “Clover” and “HIMster II” at the Helmholtz-Institut Mainz and “Mogon II” at JGU Mainz. We are grateful to our colleagues in the CLS initiative for sharing ensembles.
\newpage
\bibliographystyle{JHEP}
\normalsize
\bibliography{references}



\end{document}